\documentclass[a4paper,conference]{IEEEtran}
\IEEEoverridecommandlockouts

\usepackage[left=1.57cm,right=1.57cm,top=0.95cm,bottom=2.54cm]{geometry}
\usepackage{array,multirow}
\usepackage{booktabs} 
\usepackage{graphicx}
\usepackage{caption}
\usepackage{subcaption}
\usepackage{amssymb}
\usepackage{threeparttable}
\usepackage{booktabs}
\usepackage{cite}
\usepackage{bm}
\usepackage{amsmath,amssymb,amsfonts}
\usepackage{algorithmic}
\usepackage{graphicx}
\usepackage{mathtools}
\usepackage{textcomp}
\usepackage{xcolor}
\usepackage[hyphens]{url}
\usepackage[draft]{hyperref}
\usepackage{xcolor}

\usepackage[capitalize]{cleveref}
\crefname{section}{Sec.}{Secs.}

\def\BibTeX{{\rm B\kern-.05em{\sc i\kern-.025em b}\kern-.08em
    T\kern-.1667em\lower.7ex\hbox{E}\kern-.125emX}}

\usepackage[acronyms,nonumberlist,nopostdot,nomain,nogroupskip]{glossaries}
\newacronym{3gpp}{3GPP}{3rd Generation Partnership Project}
\newacronym{5g}{5G}{5th generation}
\newacronym{5gc}{5GC}{5G Core}
\newacronym{adc}{ADC}{Analog to Digital Converter}
\newacronym{afbw}{AFBW}{Average Fading Bandwidth}
\newacronym{aimd}{AIMD}{Additive Increase Multiplicative Decrease}
\newacronym{am}{AM}{Acknowledged Mode}
\newacronym{amc}{AMC}{Adaptive Modulation and Coding}
\newacronym[firstplural=Angles of Arrival (AoAs)]{aoa}{AoA}{Angle of Arrival}
\newacronym[firstplural=Angles of Departure (AoDs)]{aod}{AoD}{Angle of Departure}
\newacronym{ap}{AP}{Access Point}
\newacronym{app}{APP}{Application Layer}
\newacronym{aqm}{AQM}{Active Queue Management}
\newacronym{awgn}{AGWN}{Additive White Gaussian Noise}
\newacronym{balia}{BALIA}{Balanced Link Adaptation}
\newacronym{bdp}{BDP}{Bandwidth-Delay Product}
\newacronym{ber}{BER}{Bit Error Rate}
\newacronym{bf}{BF}{Beamforming}
\newacronym{cad}{CAD}{Computer-Aided Design}
\newacronym{cbr}{CBR}{Constant Bit Rate}
\newacronym{cc}{CC}{Congestion Control}
\newacronym{cdf}{CDF}{Cumulative Distribution Function}
\newacronym{ci}{CI}{Confidence Interval}
\newacronym{cicd}{CI/CD}{Continuous Integration and Developement}
\newacronym{cir}{CIR}{Channel Impulse Response}
\newacronym{cn}{CN}{Core Network}
\newacronym{cp}{CP}{Control Plane}
\newacronym{cqi}{CQI}{Channel Quality Information}
\newacronym{crs}{CRS}{Cell Reference Signal}
\newacronym{csirs}{CSI-RS}{Channel State Information - Reference Signal}
\newacronym{dc}{DC}{Dual Connectivity}
\newacronym{dce}{DCE}{Direct Code Execution}
\newacronym{dci}{DCI}{Downlink Control Information}
\newacronym{dked}{DKED}{Double Knife-Edge Diffraction}
\newacronym{dl}{DL}{Downlink}
\newacronym{dmr}{DMR}{Deadline Miss Ratio}
\newacronym{dmrs}{DMRS}{DeModulation Reference Signal}
\newacronym{dray}{D-Ray}{Deterministic Ray}
\newacronym{e2e}{E2E}{End-to-End}
\newacronym{ecn}{ECN}{Explicit Congestion Notification}
\newacronym{ecdf}{ECDF}{Empirical Cumulative Distribution Function}
\newacronym{edf}{EDF}{Earliest Deadline First}
\newacronym{em}{EM}{electromagnetic}
\newacronym{enb}{eNB}{evolved Node Base}
\newacronym{endc}{EN-DC}{E-UTRAN-\gls{nr} \gls{dc}}
\newacronym{epc}{EPC}{Evolved Packet Core}
\newacronym{es}{ES}{Edge Server}
\newacronym{fdd}{FDD}{Frequency Division Duplexing}
\newacronym{fdma}{FDMA}{Frequency Division Multiple Access}
\newacronym{fray}{F-Ray}{Flashing Ray}
\newacronym{fs}{FS}{Fast Switching}
\newacronym{ftp}{FTP}{File Transfer Protocol}
\newacronym{gmm}{GMM}{Gaussian Mixture Model}
\newacronym{gnb}{gNB}{Next Generation Node Base}
\newacronym{harq}{HARQ}{Hybrid Automatic Repeat reQuest}
\newacronym{hetnet}{HetNet}{Heterogeneous Network}
\newacronym{hh}{HH}{Hard Handover}
\newacronym{hol}{HOL}{Head-of-Line}
\newacronym{hqf}{HQF}{Highest-quality-first}
\newacronym{ia}{IA}{Initial Access}
\newacronym{iab}{IAB}{Integrated Access and Backhaul}
\newacronym{ieee}{IEEE}{Institute of Electrical and Electronics Engineers}
\newacronym{imt}{IMT}{International Mobile Telecommunication}
\newacronym{inr}{INR}{Interference to Noise Ratio}
\newacronym{iot}{IoT}{Internet of Things}
\newacronym{itu}{ITU}{International Telecommunication Union}
\newacronym{ked}{KED}{Knife-Edge Diffraction}
\newacronym{kpi}{KPI}{Key Performance Indicator}
\newacronym{ks}{KS}{Kolmogorov–Smirnov}
\newacronym{lcf}{LCF}{Level Crossing Frequency}
\newacronym{lcr}{LCR}{Level Crossing Rate}
\newacronym{los}{LoS}{Line-of-Sight}
\newacronym{lte}{LTE}{Long Term Evolution}
\newacronym{m2m}{M2M}{Machine to Machine}
\newacronym{mac}{MAC}{Medium Access Control}
\newacronym{mc}{MC}{Multi-Connectivity}
\newacronym{mcs}{MCS}{Modulation and Coding Scheme}
\newacronym{mec}{MEC}{Mobile Edge Cloud}
\newacronym{metis}{METIS}{Mobile and wireless communications Enablers for the Twenty-twenty Information Society}
\newacronym{mi}{MI}{Mutual Information}
\newacronym{mib}{MIB}{Master Information Block}
\newacronym{mimo}{MIMO}{Multiple Input, Multiple Output}
\newacronym{mlr}{MLR}{Maximum-local-rate}
\newacronym[plural=\gls{mme}s,firstplural=Mobility Management Entities (MMEs)]{mme}{MME}{Mobility Management Entity}
\newacronym{mmwave}{mmWave}{millimeter wave}
\newacronym{moi}{MoI}{Method of Images}
\newacronym{mpc}{MPC}{Multi Path Component}
\newacronym{mptcp}{MPTCP}{Multipath TCP}
\newacronym{mr}{MR}{Maximum Rate}
\newacronym{mrdc}{MR-DC}{Multi \gls{rat} \gls{dc}}
\newacronym{mss}{MSS}{Maximum Segment Size}
\newacronym{mtd}{MTD}{Machine-Type Device}
\newacronym{mtu}{MTU}{Maximum Transmission Unit}
\newacronym{nfv}{NFV}{Network Function Virtualization}
\newacronym{nist}{NIST}{National Institute of Standards and Technology}
\newacronym{nlos}{NLoS}{Non-Line-of-Sight}
\newacronym{nr}{NR}{New Radio}
\newacronym{nrmse}{NRMSE}{Normalized Root Mean Square Error}
\newacronym{ns3}{ns-3}{Network Simulator 3}
\newacronym{nsa}{NSA}{Non Stand Alone}
\newacronym{o2i}{O2I}{Outdoor-to-Indoor}
\newacronym{ofdm}{OFDM}{Orthogonal Frequency Division Multiplexing}
\newacronym{osm}{OSM}{Open Street Map}
\newacronym{pa}{PA}{Position-aware}
\newacronym{pbch}{PBCH}{Physical Broadcast Channel}
\newacronym{pdcch}{PDCCH}{Physical Downlonk Control Channel}
\newacronym{pdcp}{PDCP}{Packet Data Convergence Protocol}
\newacronym{pdsch}{PDSCH}{Physical Downlink Shared Channel}
\newacronym{pdu}{PDU}{Packet Data Unit}
\newacronym{per}{PER}{Packet Error Rate}
\newacronym{pec}{PEC}{Perfect Electrical Conductor}
\newacronym{pf}{PF}{Proportional Fair}
\newacronym{pgw}{PGW}{Packet Gateway}
\newacronym{phy}{PHY}{Physical}
\newacronym{pl}{PL}{Path Loss}
\newacronym{ppp}{PPP}{Poisson Point Process}
\newacronym{prb}{PRB}{Physical Resource Block}
\newacronym{pss}{PSS}{Primary Synchronization Signal}
\newacronym{pucch}{PUCCH}{Physical Uplink Control Channel}
\newacronym{pusch}{PUSCH}{Physical Uplink Shared Channel}
\newacronym{qd}{QD}{Quasi Deterministic}
\newacronym{qoe}{QoE}{Quality of Experience}
\newacronym{qos}{QoS}{Quality of Service}
\newacronym{rach}{RACH}{Random Access Channel}
\newacronym{ran}{RAN}{Radio Access Network}
\newacronym[firstplural=Radio Access Technologies (RATs)]{rat}{RAT}{Radio Access Technology}
\newacronym{rcs}{RCS}{Radar Cross-Section}
\newacronym{red}{RED}{Random Early Detection}
\newacronym{rf}{RF}{Radio Frequency}
\newacronym{rlc}{RLC}{Radio Link Control}
\newacronym{rlf}{RLF}{Radio Link Failure}
\newacronym{rr}{RR}{Round Robin}
\newacronym{rray}{R-Ray}{Random Ray}
\newacronym{rrc}{RRC}{Radio Resource Control}
\newacronym{rrm}{RRM}{Radio Resource Management}
\newacronym{rs}{RS}{Remote Server}
\newacronym{rsrp}{RSRP}{Reference Signal Received Power}
\newacronym{rsrq}{RSRQ}{Reference Signal Received Quality}
\newacronym{rss}{RSS}{Received Signal Strength}
\newacronym{rssi}{RSSI}{Received Signal Strength Indicator}
\newacronym{rt}{RT}{ray tracer}
\newacronym{rtt}{RTT}{Round Trip Time}
\newacronym{rw}{RW}{Receive Window}
\newacronym{rx}{RX}{Receiver}
\newacronym{sa}{SA}{standalone}
\newacronym{sack}{SACK}{Selective Acknowledgment}
\newacronym{sap}{SAP}{Service Access Point}
\newacronym{sbr}{SBR}{Shooting and Bouncing Rays}
\newacronym{sch}{SCH}{Secondary Cell Handover}
\newacronym{scm}{SCM}{Spatial Channel Model}
\newacronym{scoot}{SCOOT}{Split Cycle Offset Optimization Technique}
\newacronym{sdma}{SDMA}{Spatial Division Multiple Access}
\newacronym{sf}{SF}{Shadow Fading}
\newacronym{sked}{SKED}{Single Knife-Edge Diffraction}
\newacronym{si}{SI}{Study Item}
\newacronym{sib}{SIB}{Secondary Information Block}
\newacronym{sinr}{SINR}{Signal-to-Interference-plus-Noise Ratio}
\newacronym{sir}{SIR}{Signal-to-Interference Ratio}
\newacronym{sm}{SM}{Saturation Mode}
\newacronym{snr}{SNR}{Signal-to-Noise Ratio}
\newacronym{son}{SON}{Self-Organizing Network}
\newacronym{srs}{SRS}{Sounding Reference Signal}
\newacronym{ss}{SS}{Synchronization Signal}
\newacronym{sss}{SSS}{Secondary Synchronization Signal}
\newacronym{sta}{STA}{Station}
\newacronym{stl}{STL}{Standard Triangle Language}
\newacronym{svd}{SVD}{Singular Value Decomposition}
\newacronym{tb}{TB}{Transport Block}
\newacronym{tcp}{TCP}{Transmission Control Protocol}
\newacronym{udp}{UDP}{User Datagram Protocol}
\newacronym{tdd}{TDD}{Time Division Duplexing}
\newacronym{tdma}{TDMA}{Time Division Multiple Access}
\newacronym{tfl}{TfL}{Transport for London}
\newacronym{tgad}{TGad}{Task Group ad}
\newacronym{tgay}{TGay}{Task Group ay}
\newacronym{tm}{TM}{Transparent Mode}
\newacronym{trp}{TRP}{Transmitter Receiver Pair}
\newacronym{tti}{TTI}{Transmission Time Interval}
\newacronym{ttt}{TTT}{Time-to-Trigger}
\newacronym{tx}{TX}{Transmitter}
\newacronym{ue}{UE}{User Equipment}
\newacronym{ul}{UL}{Uplink}
\newacronym{um}{UM}{Unacknowledged Mode}
\newacronym{uma}{UMa}{Urban Macro}
\newacronym{uml}{UML}{Unified Modeling Language}
\newacronym{utc}{UTC}{Urban Traffic Control}
\newacronym{utd}{UTD}{Uniform Theory of Diffraction}
\newacronym{ven}{VENERIS}{Vehicular Networks Simulator with Realistic Physics}
\newacronym{vm}{VM}{Virtual Machine}
\newacronym{wbf}{WBF}{Wired Bias Function}
\newacronym{wf}{WF}{Wired-first}
\newacronym{wifi}{Wi-Fi}{Wireless Fidelity}
\newacronym{wigig}{WiGig}{Wireless Gigabit}
\newacronym{wlan}{WLAN}{Wireless Local Area Network}
\newacronym{xpr}{XPR}{Cross Polarization Ratio}

\usepackage{tikz}
\usepackage{pgfplots}
\usepackage{tikzscale}
\usepackage{tikz-qtree}

\pgfplotsset{compat=newest}
\pgfplotsset{plot coordinates/math parser=false}
\pgfplotsset{every axis/.append style={
                    label style={font=\scriptsize},
                    tick label style={font=\scriptsize},
                    legend style={font=\scriptsize}
                    }}
\usetikzlibrary{plotmarks,shapes,patterns,decorations.pathreplacing,backgrounds,calc,arrows,arrows.meta,spy,matrix,shadows,trees,positioning,fit}
\usepgfplotslibrary{patchplots,groupplots}

\tikzstyle{startstop} = [rectangle, rounded corners, minimum width=2cm, minimum height=0.5cm,text centered, draw=black]
\tikzstyle{io} = [trapezium, trapezium left angle=70, trapezium right angle=110, minimum width=3cm, minimum height=1cm, text centered, draw=black]
\tikzstyle{process} = [rectangle, minimum width=2cm, minimum height=0.5cm, text centered, draw=black, alignb=center]
\tikzstyle{decision} = [ellipse, minimum width=2cm, minimum height=1cm, text centered, draw=black]
\tikzstyle{arrow} = [thick,<->,>=stealth]
\tikzstyle{line} = [thick,>=stealth]
\tikzstyle{darrow} = [thick,<->,>=stealth,dashed]
\tikzstyle{sarrow} = [thick,->,>=stealth]
\tikzstyle{larrow} = [line width=0.1mm,dashdotted,->,>=stealth]

\pgfkeys{/pgf/number format/.cd,1000 sep={\,}} % thousand separator: space

\makeatletter
\def\grd@save@target#1{%
  \def\grd@target{#1}}
\def\grd@save@start#1{%
  \def\grd@start{#1}}
\tikzset{
  grid with coordinates/.style={
    to path={%
      \pgfextra{%
        \edef\grd@@target{(\tikztotarget)}%
        \tikz@scan@one@point\grd@save@target\grd@@target\relax
        \edef\grd@@start{(\tikztostart)}%
        \tikz@scan@one@point\grd@save@start\grd@@start\relax
        \draw[minor help lines] (\tikztostart) grid (\tikztotarget);
        \draw[major help lines] (\tikztostart) grid (\tikztotarget);
        \grd@start
        \pgfmathsetmacro{\grd@xa}{\the\pgf@x/1cm}
        \pgfmathsetmacro{\grd@ya}{\the\pgf@y/1cm}
        \grd@target
        \pgfmathsetmacro{\grd@xb}{\the\pgf@x/1cm}
        \pgfmathsetmacro{\grd@yb}{\the\pgf@y/1cm}
        \pgfmathsetmacro{\grd@xc}{\grd@xa + \pgfkeysvalueof{/tikz/grid with coordinates/major step x}}
        \pgfmathsetmacro{\grd@yc}{\grd@ya + \pgfkeysvalueof{/tikz/grid with coordinates/major step y}}
        \foreach \x in {\grd@xa,\grd@xc,...,\grd@xb}
        \node[anchor=north] at (\x,\grd@ya) {\pgfmathprintnumber{\x}};
        \foreach \y in {\grd@ya,\grd@yc,...,\grd@yb}
        \node[anchor=east] at (\grd@xa,\y) {\pgfmathprintnumber{\y}};
      }
    }
  },
  minor help lines/.style={
    help lines,
    gray,
    line cap =round,
    xstep=\pgfkeysvalueof{/tikz/grid with coordinates/minor step x},
    ystep=\pgfkeysvalueof{/tikz/grid with coordinates/minor step y}
  },
  major help lines/.style={
    help lines,
    line cap =round,
    line width=\pgfkeysvalueof{/tikz/grid with coordinates/major line width},
    xstep=\pgfkeysvalueof{/tikz/grid with coordinates/major step x},
    ystep=\pgfkeysvalueof{/tikz/grid with coordinates/major step y}
  },
  grid with coordinates/.cd,
  minor step x/.initial=.5,
  minor step y/.initial=.2,
  major step x/.initial=1,
  major step y/.initial=1,
  major line width/.initial=1pt,
}
\makeatother

\newlength\fheight
\newlength\fwidth

\def \dfwidth{0.4\linewidth}
\def \dfheight{0.45\linewidth}
\def \tfwidth{0.25\linewidth}

\def \qfwidth{0.23\linewidth}
\def \qfheight {0.5\linewidth}

\begin{document}

\title{\vskip 4.5mm An Open Framework to Model Diffraction by Dynamic Blockers in Millimeter Wave Simulations}

% Alternative titles:
% - 

\author{\IEEEauthorblockN{Paolo Testolina, Mattia Lecci, Alessandro Traspadini, and Michele Zorzi
\thanks{This work was partially supported by the National Institute of Standards and Technology (NIST) under award no. 60NANB20D082.
The work of M. Lecci and P. Testolina was supported by Fondazione CaRiPaRo under grants ``Dottorati di Ricerca'' 2018 and 2019.}}
    
\IEEEauthorblockA{Department of Information Engineering, University of Padova, Italy\\
E-mail: 
\texttt{\{testolina,leccimat,traspadini,zorzi\}@dei.unipd.it}}
}

\IEEEoverridecommandlockouts
\newcommand\copyrightnotice{%
	\begin{tikzpicture}[remember picture,overlay]
	\node[anchor=south,yshift=10pt] at (current page.south) {\fbox{\parbox{\dimexpr\textwidth-\fboxsep-\fboxrule\relax}{
				\footnotesize \textcopyright 2022 IEEE. Personal use of this material is permitted.
				Permission from IEEE must be obtained for all other uses, in any current or future media,
				including reprinting/republishing this material for advertising or promotional purposes,
				creating new collective works, for resale or redistribution to servers or lists,
				or reuse of any copyrighted component of this work in other works.}}};
	\end{tikzpicture}
}

\maketitle

\copyrightnotice

\begin{abstract}
	The \gls{mmwave} band will be exploited to address the growing demand for high data rates and low latency.
	The higher frequencies, however, are prone to limitations on the propagation of the signal in the environment.
	Thus, highly directional beamforming is needed to increase the antenna gain.
	Another crucial problem of the \gls{mmwave} frequencies is their vulnerability to blockage by physical obstacles.
	To this aim, we studied the problem of modeling the impact of second-order effects on \gls{mmwave} channels, specifically the susceptibility of the \gls{mmwave} signals to physical blockers. With respect to existing works on this topic, our project focuses on scenarios where \glspl{mmwave} interact with multiple, dynamic blockers. Our open source software includes diffraction-based blockage models and interfaces directly with an open source \gls{rf} ray-tracing software.
\end{abstract}


\begin{tikzpicture}[remember picture,overlay]
\node[anchor=north,yshift=-10pt] at (current page.north) {\parbox{\dimexpr\textwidth-\fboxsep-\fboxrule\relax}{
		\centering\footnotesize This paper has been accepted for presentation at the 20th Mediterranean Communication and Computer Networking Conference. \textcopyright 2022 IEEE.\\
Please cite it as:
P. Testolina, M. Lecci, A. Traspadini, and M. Zorzi, “An Open Framework to Model Diffraction by Dynamic Blockers in Millimeter Wave Simulations,” in IEEE 20th Mediterranean Communication and Computer Networking Conference (MedComNet), Paphos, Cyprus, Jun. 2022.
}};
\end{tikzpicture}

\glsresetall

\section{Introduction}
\label{sec:introduction}

The \gls{mmwave} frequencies feature large chunks of untapped bandwidth that can increase the data rate provided to the end users, and the small wavelength enables the design of antenna arrays with tens of elements in a small form factor to support
beamformed transmissions.
While these promising characteristics make the \gls{mmwave} technology able to meet the requirements of \gls{5g} cellular systems and Wi-Fi networks~\cite{rappaport2013millimeter,rangan2017potentials}, there are several concerns regarding the propagation characteristics at these frequencies that justify a more accurate study and the need for new channel models.
First, the high propagation loss limits the coverage region of \gls{mmwave} base stations, although large antenna arrays and complex beamforming techniques can mitigate the problem.
Second, at \glspl{mmwave}, the increased diffraction loss results in deep shadow regions, thus further degrading the communications performance~\cite{dengMmwaveDiffraction,hemadeh2018millimeter}.
Furthermore, \gls{mmwave} signals can be easily blocked by obstacles (e.g., vehicles, buildings, vegetation, human bodies), which may prevent direct \gls{los} communications. 
These (often unpredictable) propagation components make it imperative to accurately model the dynamics of the surrounding environment, and to design communication protocols, and especially beam-tracking algorithms, taking into account these disruptive events.

In this context, ray tracing has emerged as an essential tool to model the \gls{mmwave} channel~\cite{testolina2020simplified}, especially, but not limited to, when detailed link-level protocols need to be simulated.
With respect to stochastic channel models, \glspl{rt} exploit a digital reconstruction of the environment to achieve a greater degree of accuracy, at the cost of additional computational load~\cite{lecci2020accuracy}.
Generally, the digital model includes the static elements of the scenario, e.g., the floor, ceiling and walls, and the tables, screens and other objects, depending on its level of detail.
Its design is not trivial, as it requires the use of a \gls{cad} software, a complex and time-consuming task.
Furthermore, dynamic elements, e.g., moving humans and vehicles, are not generally considered, as their movements can not be included in a single \gls{cad} file.
Thus, these elements are often overlooked in \gls{rt} simulations, despite the significant role they play in the propagation of the signal~\cite{gentile2021HumanPresence,macCartney2016humanBlockage}.
Furthermore, the study of the impact of blockers on the network performance has been limited to ad hoc scenarios, often considering a single blocker and only at the PHY layer.
On the contrary, their effect on large-scale, high-level network simulations has not yet been fully characterized, due to the complexity of designing the scenario and modeling the diffraction when considering moving obstacles.

In this work, we present the \textit{Blockage Manager}\footnote{The \textit{Blockage Manager} is available at \url{https://github.com/signetlabdei/rt-blockage-manager}.}, a novel open-source software to model dynamic blockers in \gls{rt} simulations.
% Specifically, we consider the {qd-realization} software, an open source \gls{rt} simulator by the \gls{nist}. \ml{it looks like it need qd-realization to survive and it doesn't make sense otherwise. it is like that for the moment, but it can be extended}
% Currently, the {qd-realization} software does not support dynamic blockers, and their introduction would require either significant modifications to the \gls{rt} code to introduce the notion of ``Obstacle'' and its interactions with the rays (e.g., obstruction, diffraction, etc), or creating dynamic \gls{cad} scenarios, accounting for the movement of the dynamic elements in each time step.
The application was designed to post-process information typically obtained from \gls{rt} software, offering the user a simple yet powerful interface to introduce blockage models on top of them, making their simulations more realistic and dynamic.
Starting from a pre-processed simulation allows the user to later add as many obstacles as desired, with custom mobility and settings, without the need to run an entirely new ray-tracing simulation from scratch, which can take a significant amount of time.
The software has been designed to present as simple an API as possible, so that the user only has to create a high-level description of the desired obstacles, while still having some control on fine-tuning parameters if desired, and leaving all the complexity to the simulator. 

The aim of this work is to describe an open-source software able to (i) interface directly with \textit{qd-realization}~\cite{qd-realization}, an open-source RF ray-tracing software, importing and exporting channel traces with a single line of code, (ii) process pre-computed \gls{rt} simulations, adding multiple mobile obstacles on top of them, (iii) support already implemented blockage models, with varying degrees of complexity and accuracy.
Furthermore, we showcase some basic network scenarios processed with our \textit{Blockage Manager}, describing the obtained results and discussing future works that will enable it to closely mimic real channel behaviors in the presence of obstacles, allowing the community to better study and design link-level protocols for \gls{mmwave} communications.

In the remainder of the paper, we first report the state of the art on diffraction modeling (\cref{sec:soa}).
Then, \cref{sec:software_architecture} offers a brief overview of the framework architecture, giving the essential details to understand its main components.
In \cref{sec:blockage_models}, we present the blockage models that are currently implemented in the framework.
Finally, the results in \cref{sec:results} serve the double purpose of showcasing the usage of the \textit{Blockage Manager} and of highlighting the impact of dynamic blockage in \gls{mmwave} networks, and \cref{sec:conclusions} presents some concluding remarks and future work directions.

\section{State of the Art}
\label{sec:soa}

Signals at \gls{mmwave} frequencies are prone to limitations on their propagation in the environment.
Therefore, understanding and precisely characterizing the interaction between radio waves and the surroundings is fundamental to characterize communications in different scenarios.

This reason has led to different modeling approaches for the \gls{mmwave} channel, which have various degrees of complexity and accuracy, and can be applied to different contexts and evaluations~\cite{lecci2020accuracy}.
Thus, channel models can be summarized in: analytical, stochastic, and \gls{qd}.

\paragraph*{Analytical channel models} These models generally offer a simplified representation of the channel, based on propagation loss and a random variable representing fading.
This kind of model has a limited accuracy and can be used to characterize communications in a generic environment~\cite{bai2015coverage,andrews2017modeling,ferrand2016trends}, without accounting for the features of specific scenarios and their interaction with the elements typical of \gls{mmwave} propagation (e.g., realistic antenna arrays and beamforming).

\paragraph*{Stochastic channel models} They derive the entries of the channel matrix from a set of random distributions, whose parameters are determined by statistical fits on channel measurements for a generic scenario (e.g., a common rural or urban environment).
Their speed and their stochastic nature allow to easily generate the radio channels for generic, large-scale scenarios.
Models of this type, e.g., the one proposed in 3GPP TR~38.901~\cite{3gpp.38.901}, have been used in the performance evaluations of \gls{mmwave} networks~\cite{gapeyenko2018analytical,zugno2020implementation}.

\paragraph*{Quasi-Deterministic channel models} \gls{qd} channels, instead, can accurately model the interactions of the \gls{mmwave} signal with a specific environment~\cite{lecci2020qd,degliesposti2014rt,mckown91rt}.
However, since the \glspl{mpc} are generated by the interaction between the transmitted signal and the elements of a given environment, they can offer improved accuracy at the cost of a higher processing load.

Including the effect of blockage on the channel model is a non-trivial task.
The high-level, generic nature of the analytical models makes it difficult to model the blockage by multiple, possibly moving obstacles (e.g., people, vehicles, foliage, etc.).
Similarly, since stochastic channel models do not require a model of the propagation environment, it is difficult to define a temporally and spatially consistent behavior of the blockers.
Furthermore, modeling specific mobility patterns does not agree well with the stochastic nature of the channel, and can decrease the advantages of these kinds of models (i.e., their being lightweight and easy to use).
Finally, as mentioned in \cref{sec:introduction}, placing (possibly dynamic) blockers in an \gls{rt} simulation can also be challenging, thus motivating this work.
In the remainder of this section, we present the existing models in the literature that deal with this problem.
Namely, we first report the main works on blockage at the channel modeling level, and then its modeling at the higher layers of the network stack.

The authors of~\cite{schuler08scatteringCenters} analyzed the scattering of a vehicle and showed that \gls{rt} simulations of complex objects allow the derivation of simplified scattering models.
This is fundamental, since only by reducing the complexity of large objects does it become possible to simulate realistic scenarios.
Their results show that, rather than introducing a complex model with a high polygon count, it is possible to reduce the complexity of simulations involving cars by only considering a few key points producing scattering, e.g., wing mirrors.

Another simplification was described also for human blockage in~\cite{ghaddar07conductingCylinder}, where the authors analyzed their measurements taken at 10.5~GHz and concluded that a metal cylinder affects the radio channel similarly to a human body.
This result justifies the usage of such a simple model in practical simulations rather than modeling a detailed person, a significantly harder task to perform with a correspondingly much higher computational cost.

The 3GPP also proposes a map-based hybrid channel model~\cite{3gpp.38.901} using~\cite{metisChannelModels_D1.4} for the deterministic component of the channel, and adding a random component to model clusters of \glspl{mpc}.
It also proposes the usage of two blockage models, to better represent a dynamic and realistic channel.
In the stochastic blockage model (Blockage model A), random rectangular regions are blocked, considering the temporal and spatial correlation for completeness.
The geometric blockage model (Blockage model B) deploys rectangular screens, specifying their relative dimensions and mobility patterns.
In both cases, a \gls{ked} at four edges is defined, using the $\arctan$ approximation, described in~\cref{sub:metis}, for the diffraction loss.

A low-complexity enhancement of this approximation was studied in~\cite{medbo2013channel} using a Fourier-based model for human obstacles.
Its authors compared their simulations with measurements at 2.44~GHz in an indoor \gls{nlos} scenario and showed that their approximation produces shadowing loss closer to the Fresnel formulas for \gls{dked} than the simpler $\arctan$ model.

Besides, the validity of the \gls{ked} model was proved for narrow obstacles also using complex specific solvers for Maxwell's equations as described in~\cite{casciato01thesis}.

A series of modifications and criteria to improve the accuracy of the 3GPP model are provided in \cite{prado21enhance}.
It proposed, in case of multiple blockers in close proximity, to consider them as a single blocker with increased size.
The authors also suggest a way to increase the accuracy when the transmitted beam does not cover the entire obstacle surface, for instance, when highly directional antennas at a relative short distance are employed.
In such cases, the idea consists of taking into consideration the transmitted beam pattern to choose the position of the diffraction points on the surface of the obstacles.
A measurement campaign was carried out, and its results showed that with these modifications the loss prediction provided by the 3GPP model can be improved.

The literature also includes some studies analyzing blockage experimentally~\cite{gentile2021HumanPresence,macCartney2016humanBlockage,slezak18under}.
The authors in~\cite{gentile2021HumanPresence} analyzed accurate measurements of human body blockage at 60~GHz with a person that traversed a linear trajectory perpendicular to the direct path between the \gls{tx} and the \gls{rx}. 
They derived a diffraction model and compared it with their measurements.
Eventually, they showed that when a human obstacle is close to the path of the direct ray, it also creates a strong reflected ray, thus acting as a 2-ray propagation model.

A similar scenario was studied in \cite{macCartney2016humanBlockage} at 73~GHz.
Its analysis showed that the use of directive antennas has a crucial impact on human blockage measurement results, especially when the blocker was close either to the \gls{tx} or to the \gls{rx}.

Furthermore,~\cite{jacob11humanBlockage} explores the impact of human blockage on the channel model specifically proposed for IEEE~802.11ad~\cite{tgad_channel_model}, based on measurements at 60~GHz.
Its authors analyzed ray tracing simulations in a human blockage scenario considering also diffraction effects.
They found that the impact of human blockage is twofold.
First of all, the \gls{snr} at the \gls{rx} decreases due to the additional attenuation, even if smart antennas are employed.
On the other hand, the delay spread and the frequency selectivity in general increase, because the channel conditions change from \gls{los} to \gls{nlos}.
Both effects deteriorate the link, causing a higher bit error probability during communications.

The modern literature not only provides examples of works studying blockage at mmWaves from a physical measurement and modeling perspective, but also recognizes it as an issue that mmWave communication protocols need to address.
In fact, while high frequencies are in principle capable of delivering the exceptional performance that they promise, they are also heavily affected by bad channel conditions, making the system much less stable thus worsening the \gls{qoe}.

An analysis of blockage was performed in~\cite{slezak18under} to evaluate the effect of blockers on an \gls{e2e} application.
The authors used a phased array system that measured the channel in multiple directions, in order to fully understand the path diversity.

Different analyses on \gls{tcp} \cite{zhang2019will,polese17mobilitymanagement,polese17milliproxy,polese17tcpin5g}, studying the behavior of this protocol at \gls{mmwave} frequencies, employed obstacles to simulate \gls{nlos} scenarios.
Indeed, the extreme variability of the signal quality over mmWave links yields either degraded \gls{tcp} goodput and very low utilization of the resources at \gls{mmwave} frequencies, or, in the presence of link-layer retransmissions, high goodput at the price of high latency.
Therefore, especially at \glspl{mmwave}, the effect of obstacles is crucial to understand the actual condition of the channel.

As blockage is so impactful, especially on \gls{tcp}-based communications, some works have proposed and tested ways of relying on multiple simultaneous connections, on both high and low frequencies, to balance resilience and raw throughput. Protocols such as \gls{mptcp} have been analyzed~\cite{polese17tcp,polese17tcpin5g,mezzavilla2018end,ren2021tcpMmwave,Imran22mptcpSmartphones,lee19mptcp}, showing their advantages when, for example, buildings block the direct mmWave path or fast mobility cannot be fully supported by the mmWave network.

In \cite{polese17mobilitymanagement}, the authors assessed the goodput of a scenario in which obstacles of different sizes are placed in the area between the \gls{tx} and the \gls{rx}.
The application layer simulated a file transfer with full buffer and the authors noticed that the number of obstacles plays a major role in the achievable goodput.
With fewer obstacles, there is a higher probability of having a \gls{los} channel and this has a crucial impact on the data rate available at the physical layer.

Thus, the use of more accurate channel models, modeling for instance the diffraction effect of obstacles and considering appropriately the presence of multiple blockers, can provide more precise results.
For example, even in a simple human blockage scenario where an obstacle moves perpendicularly to the line between the \gls{tx} and the \gls{rx}, the loss is higher when the obstacle crosses the \gls{los} closer to one of the nodes.
This phenomenon, shown in our simulations, can be described only by taking into account the diffraction effect of the obstacles.
Thus, it can be easily understood that in more complex scenarios with many obstacles this effect has a critical impact on results.
This is the reason why we implemented an open source software able to (i) include diffraction-based blockage models, (ii) interface directly with an open source \gls{rf} ray-tracing software, (iii) support multiple mobile blockers.

%\ml{
%	\begin{itemize}
%		\item there is literature analyzing it experimentally (limited to extremely simple scenarios, e.g., \cite{gentile2021HumanPresence}), often focusing on modeling blockage
%		\item there is literature showing possible solutions from higher layer on how to handle a blockage event, but simple blockage models are used \cite{ford16ns3mmwave,assasa16implementation11ad,assasa17extending11ad}
%		\item standards include blockage in their channel models, both stochastic and geometric (\cite{3gpp.38.901,metisChannelModels_D1.4}, double check \cite{tgad_channel_model,tgay_channel_model})
%		\item our novelty (not sure if we should put this in the intro): providing an open source software able to (i) include diffraction-based blockage models, (ii) interface directly with an open source RF ray-tracing software, (iii) support multiple mobile blockers, (iv) ...
%	\end{itemize}
%}

\section{Software Architecture}
\label{sec:software_architecture}

% \ml{
%     \begin{itemize}
%         \item brief description of the main src files
%         \item interfaces that allow simple extension of the software (mobility models, diffraction loss, shapes, etc)
%         \item interaction with qd-realization software and pipeline description
%         \item why we chose to post-process with this software
%     \end{itemize}
% }
The \textit{Blockage Manager} was designed as a self-contained software to model the effect of blockage in \gls{rt} traces.
Doing it in post-processing allows running the computationally-demanding \gls{rt} only once, including in the simulation all the static elements with a single \gls{cad} model.
Any distribution of the dynamic elements, i.e., blockers, can then be introduced  in the static scenario using the \textit{Blockage Manager}, that can use the same \gls{rt} output as a static baseline for any blocker configuration.

The \textit{Blockage Manager} is organized in modules, each providing a set of classes and functions to handle the elements required for the simulation.
Here, we report a brief overview of the different modules, to offer some insights on the implementation and to clarify how the results presented in \cref{sec:results} were obtained.

Although geometry libraries exist in Python~\cite{simpy,shapely}, in order to meet all the simulation requirements (3D geometry with fast computation time), and to build a self-contained package, we decided to implement all the basic geometrical objects and operations from scratch.
Thus, we developed a custom \texttt{Geometry} module to handle the geometry of the ray and of the obstacles themselves, and to define consistent mobility patterns of the latter.
This module underlies most of the operations carried out during the simulation and can be optimized independently in the future, thanks to the common object interface.

The \gls{rt} traces can be imported in the \textit{Blockage Manager} using the \texttt{Scenario} interface, that defines common methods to import/export traces in the target format, as well as to access and update sets of rays between nodes.
Currently, the \texttt{Scenario} interface supports traces from the \textit{qd-realization} software, which is specifically able to handle channel traces for multiple users and timesteps.
However, the interface allows for the possibility to support different ray-tracing formats, that could be implemented in the future.
The information for each ray (delay, path gain, phase, path) is stored in a \texttt{Ray} object, that also offers a simple interface to consistently compute \glspl{aod} and \glspl{aoa}.

Then, blockers can be introduced in the scenario with the common \texttt{Obstacle} interface, that handles obstructions, diffraction, and other effects that a generic obstacle may impose over the \texttt{Ray}s of the imported \texttt{Scenario}.
Currently, a sphere, a rectangular and an \textit{orthogonal-rectangular} screens are implemented.
We defined the orthogonal-rectangular obstacle as an ideal rectangular screen that behaves as if it were orthogonal to any considered ray, when computing the interaction between the two.
As detailed in \cref{sec:blockage_models}, this artificial obstacle was introduced to meet the hypotheses of several diffraction models, that could thus be included in the software.
On the contrary, the rectangular screen can be tilted in both the azimuth and elevation directions, thus providing a more general obstacle mobility at the cost of a limited set of available diffraction models.
The movement of an obstacle during the simulation is described by a \texttt{MobilityModel}.
At each time step, the position of each obstacle is updated based on such model, thereby providing accurate and temporally-correlated mobility and making the channel temporally consistent.
The whole simulation, i.e., computing the interactions of the rays with the obstacles, and updating the positions of the obstacles at each time step, is handled by the \texttt{Environment}, that constitutes the core of the software, with just a few lines of code.

\section{Blockage Models}
\label{sec:blockage_models}

\begin{figure}[t!]
	\includegraphics[width=\columnwidth]{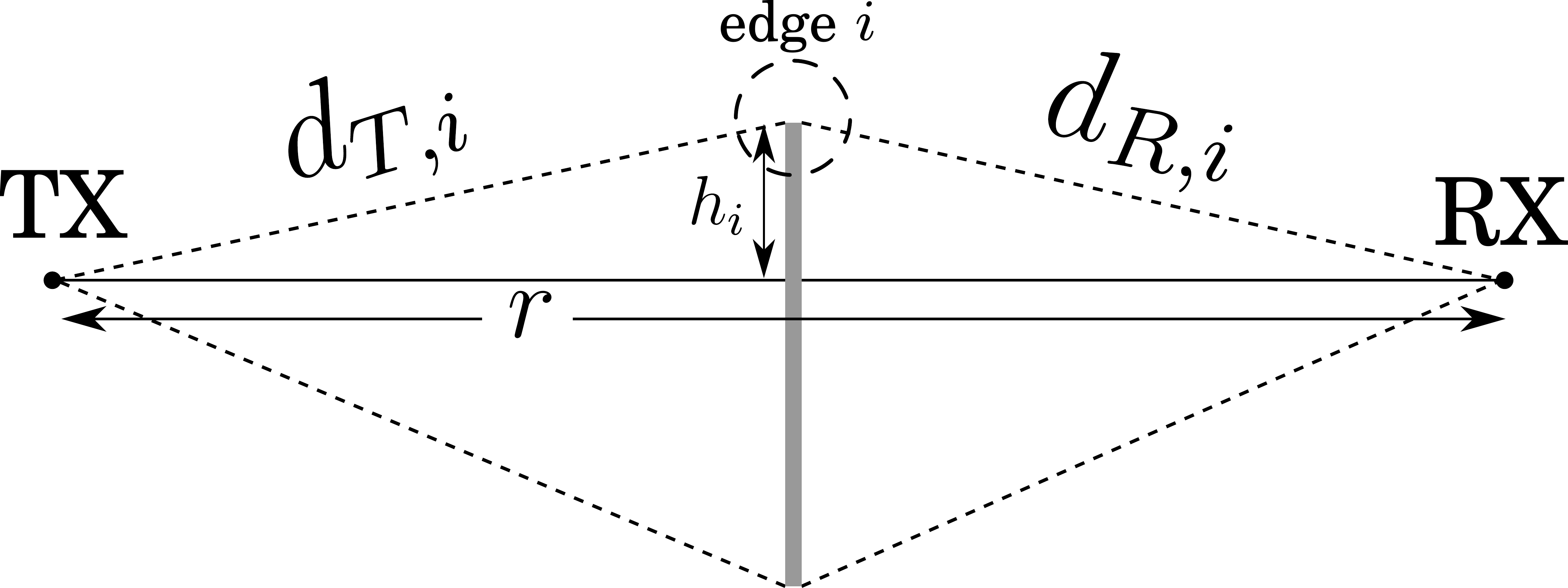}
	\caption{\gls{dked} geometry.}
	\label{fig:ob_metis}
\end{figure}

\begin{figure*}[t!]
	\hspace*{\fill}
	\begin{subfigure}[b]{\dfwidth}
		\centering
		\setlength\fwidth{\columnwidth} 
		\setlength\fheight{\dfheight}
		\input{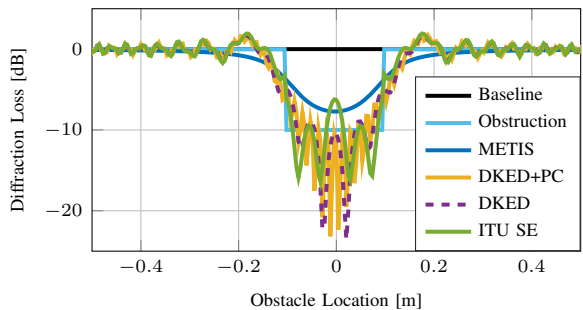}
		\caption{Models at 60 GHz}
		\label{fig:models_60ghz}
	\end{subfigure}
	\hspace*{\fill}
	\begin{subfigure}[b]{\dfwidth}
		\centering
		\setlength\fwidth{\columnwidth} 
		\setlength\fheight{\dfheight}
		\input{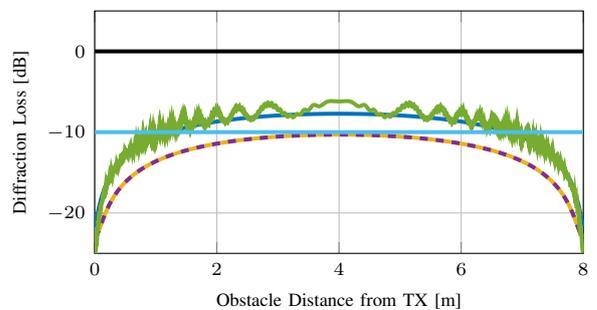}
		\caption{Different obstacle's distance}
		\label{fig:models_obstacle_distance}
	\end{subfigure}
	% \hspace*{\fill}
	% \begin{subfigure}[b]{\tfwidth}
	%   \centering
	%   \caption{Different RX distance \ml{needs some extra work}}
	%   \label{fig:models_txrx_distance}
	% \end{subfigure}
	\hspace*{\fill}
	\caption{Comparison between the implemented models using a carrier frequency of 60~GHz.}
	\label{fig:model_comparison}
\end{figure*}

\begin{figure*}[t!]
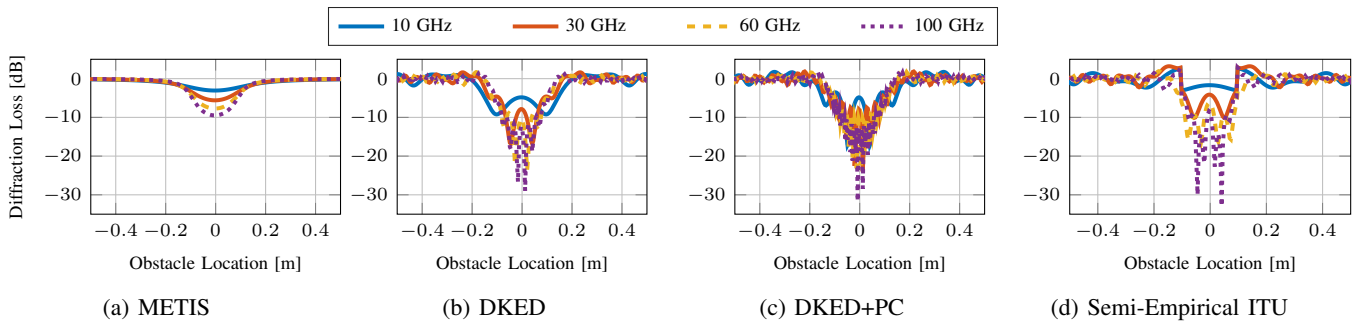

	
	\begin{subfigure}[b]{\linewidth}
		\centering
		% This file was created by matlab2tikz.
%
%The latest updates can be retrieved from
%  http://www.mathworks.com/matlabcentral/fileexchange/22022-matlab2tikz-matlab2tikz
%where you can also make suggestions and rate matlab2tikz.
%
\definecolor{mycolor1}{rgb}{0.00000,0.44700,0.74100}%
\definecolor{mycolor2}{rgb}{0.85000,0.32500,0.09800}%
\definecolor{mycolor3}{rgb}{0.92900,0.69400,0.12500}%
\definecolor{mycolor4}{rgb}{0.49400,0.18400,0.55600}%
\begin{tikzpicture}
\pgfplotsset{every tick label/.append style={font=\scriptsize}}

\begin{axis}[%
width=0,
height=0,
at={(0,0)},
scale only axis,
xmin=0,
xmax=0,
xtick={},
ymin=0,
ymax=0,
ytick={},
axis background/.style={fill=white},
legend style={legend cell align=center,
              align=center,
              draw=white!15!black,
              at={(0, 0)},
              anchor=center,
              /tikz/every even column/.append style={column sep=2em}},
legend columns=4,
]
\addplot [color=mycolor1, line width=1.5pt]
  table[row sep=crcr]{%
-1.50144	-0.0208000000000084\\
};
\addlegendentry{10 GHz}

\addplot [color=mycolor2, line width=1.5pt]
  table[row sep=crcr]{%
-1.50144	-0.0136999999999858\\
};
\addlegendentry{30 GHz}

\addplot [color=mycolor3, dashed, line width=1.5pt]
  table[row sep=crcr]{%
-1.50144	-0.01039999999999\\
};
\addlegendentry{60 GHz}

\addplot [color=mycolor4, dotted, line width=1.5pt]
  table[row sep=crcr]{%
-1.50144	-0.00840000000000174\\
};
\addlegendentry{100 GHz}

\end{axis}
\end{tikzpicture}%
	\end{subfigure}
	\\
	\hspace*{\fill}
	\begin{subfigure}[b]{\qfwidth}
		\centering
		\setlength\fwidth{\columnwidth}
		\setlength\fheight{\qfheight}
		\input{img/HumanPresence_LoS_atan_diffraction.tex}
		\caption{METIS}
		\label{fig:models_frequency_atan}
	\end{subfigure}
	\hspace*{\fill}
	\begin{subfigure}[b]{\qfwidth}
		\centering
		\setlength\fwidth{\columnwidth}
		\setlength\fheight{\qfheight}
		\input{img/HumanPresence_LoS__kunisch.tex}
		\caption{DKED}
		\label{fig:models_frequency_kunisch}
	\end{subfigure}
	\hspace*{\fill}
	\begin{subfigure}[b]{\qfwidth}
		\centering
		\setlength\fwidth{\columnwidth}
		\setlength\fheight{\qfheight}
		\input{img/HumanPresence_LoS__geom_emp.tex}
		\caption{DKED+PC}
		\label{fig:models_frequency_geom_emp}
	\end{subfigure}
	\hspace*{\fill}
	\begin{subfigure}[b]{\qfwidth}
		\centering    
		\setlength\fwidth{\columnwidth}
		\setlength\fheight{\qfheight}
		\input{img/HumanPresence_LoS_empirical_itu.tex}
		\caption{Semi-Empirical ITU}
		\label{fig:models_frequency_empirical_itu}
	\end{subfigure}
	\hspace*{\fill}
	
	\caption{Comparison of the implemented models at different frequencies.}
	\label{fig:model_comparison_frequency}
\end{figure*}

When considering the interaction between an object and a signal propagating in the free space from point $T$ to point $R$, the Fresnel Zones offer a useful model to analyze the intensity of the diffraction.
The Fresnel zones are concentric ellipses with focal points at $T$ and $R$, and radius
\begin{equation}
	r_n = \sqrt{n\lambda\frac{d_T d_R}{d_T+d_R}}
\end{equation}
where $\lambda$ is the wavelength of the transmitted signal, $d_T$ ($d_R$) is the distance between the diffraction point and the transmitter (receiver), and $n$ is the order of the Fresnel Zone.
If the cross section of the First Fresnel Zone is obstructed, diffraction becomes the dominant term in the interaction with the object, with significant impact on the communication performance.
In the \textit{Blockage Manager}, several models to describe the diffraction loss are available.
In this section, we provide a brief overview of those that were considered for the simulations described in \cref{sec:results}.

As the majority of the diffraction models is based on the Fresnel formulation, we define the complex Fresnel integral as:
\begin{equation}
	\label{eq:fresnel_int}
	F(\nu)=\int_0^\nu e^{j\frac{\pi s^2}{2}} \mathrm{d}s
\end{equation}
where $\nu$ is the Fresnel-Kirchhoff diffraction parameter:
\begin{equation}
	\nu(h) = h \sqrt{\frac{2}{\lambda}\frac{d_T + d_R}{d_T d_R}}
\end{equation}
and depends on the obstruction depth $h$.
The real and imaginary parts of $F(\nu)$ can be recalled as follows:
\begin{align*}
	F(\nu)= C(\nu)+jS(\nu)
\end{align*}
The Fresnel integral can be computed by numerical calculation, however~\cite{itu-r-p526} provides a simple way to compute an approximate result.

\subsection*{Obstruction}
\label{sub:obstruction}
When an obstacle cuts through a transmission path, the simplest obstruction model consists in applying a constant loss $L$ during the shadowing window, resulting in a sharp transition between an obstructed and an unobstructed path.
The obstruction loss is generally computed based on the absorption properties of the obstacle.
Depending on the model assumptions, an obstructed path may be completely removed from the scenario.
Its simplicity makes this model desirable in terms of computational and implementation complexity, so that it has often been chosen for end-to-end network simulations~\cite{ford16ns3mmwave,assasa16implementation11ad,assasa17extending11ad}.
Unfortunately, the unrealistic sharp drop not only yields imprecise results, but can also affect some beam-tracking algorithms, which are of the utmost importance when it comes to mmWave communication.

\subsection*{METIS}
\label{sub:metis}
The \gls{metis} channel model~\cite{metisChannelModels_D1.4} provides a simplified method to account for the diffraction contribution at \gls{mmwave} frequencies.
The screen is assumed to be vertical and perpendicularly oriented with respect to the considered ray segment in the projection from above.
The diffraction loss is modeled using a \gls{ked} model for the four edges of the screen as:
\begin{equation}
L = -20\log_{10}\left(1-(l_{h1}+l_{h2})(l_{w1}+l_{w2})\right) \quad \text{[dB]}
\end{equation}
where $l_{hi}$ and $l_{wi}$ are the \gls{sked} at edge $i$, corresponding to the height $h$ and the width $w$ of the obstacle.
The loss of each single edge is provided by:
\begin{equation}
l_{i} = \frac{\text{atan}\left(\pm\frac{\pi}{2}\sqrt{\frac{\pi \left(d_{T,i} + d_{R,i} - r \right)}{\lambda}}\right)}{\pi}
\end{equation}
where, as shown in~\cref{fig:ob_metis}, $d_{T,i}$ and $d_{R,i}$ are the distances between the nodes and edge $i$ of the
screen, and $r$ is the distance between the \gls{tx} and the \gls{rx}.
If the link is \gls{nlos}, then each contribution is taken as positive, whereas for \gls{los} condition, only the farthest edge from the link provides a positive contribution.

\subsection*{Double Knife-Edge Diffraction (DKED)}
This model~\cite{kunisch08dked} computes the \gls{sked} from the screen edges evaluating both amplitudes and phases.
Since only the lateral edges are considered for the \gls{ked}, the interaction of the obstacle with the propagating waves is that of a vertical stripe of infinite vertical extent.
The \gls{sked} for edge $i$ can be computed as:
\begin{equation}
	\label{eq:sked}
	l_i = \frac{1+j}{2} \left[\left(\frac{1}{2}-C(\nu_i)\right)-j\left(\frac{1}{2} - S(\nu_i)\right)\right]
\end{equation}
and then, this model computes the superposition of each contribution to obtain the diffraction loss:
\begin{equation}
	L = -20\log_{10}(|l_1+l_2|) \quad \text{[dB]}
\end{equation}
Besides, as the \textit{Semi-empirical ITU} model, it provides a valid solution even for non-orthogonal obstacles, and thus can be exploited for a larger variety of obstacles.

\subsection*{Double Knife-Edge Diffraction with Phase Correction (DKED+PC)}
In~\cite{kunisch08dked}, the authors present also a modified version of the \gls{dked}.
Namely, the \gls{sked} is computed for the two contributions (front and back of the body) as described by~\cref{eq:sked} for the \gls{dked} model.
Additionally, this method takes into account the phase shift of the diffracted rays path when combining the two contributions.
\begin{equation}
L = -20\log_{10}\left(\left|l_1\cdot e^{-j\frac{2\pi \Delta d_1}{\lambda}}+l_2 \cdot e^{-j\frac{2\pi \Delta d_2}{\lambda}}\right|\right) \quad \text{[dB]}
\end{equation}
where $\Delta d_i$ is the length of the $i$-th diffracted ray.
The model is presented and calibrated for ultra-wideband measurements between 4 and 10~GHz.
However, its analytical derivation is based on the Fresnel diffraction theory, that holds also for higher frequencies, as reported in \cite{gentile2021HumanPresence}.

\subsection*{ITU-R P.526-15}
The \gls{itu} provides guidelines on the modeling of the diffraction in~\cite{itu-r-p526}.
Based on the \gls{utd} and on high-precision measurements, two approximations are given for estimating the minimum and the average diffraction loss.
Furthermore, a Semi-Empirical method (ITU SE) is derived for thin rectangular screens, that exhibits the rapid fluctuations of the field strength due to the constructive and destructive interference of the diffraction from the edges.
This method can be used also for non-orthogonal obstacles.
Moreover, with reasonable accuracy, it does not require the Fresnel integral to be solved.
Both models can be applied when the wavelength is fairly small in relation to the size of the obstacles, which is the case when considering human-size blockers at mmWave frequencies.
The mathematical details of this method are reported in~\cite{itu-r-p526}.

\section{Results}
\label{sec:results}

\begin{figure*}[t!]
	\hspace*{\fill}
	\centering
	\begin{subfigure}[b]{.3\linewidth}
		%\centering
		\includegraphics[width=\columnwidth]{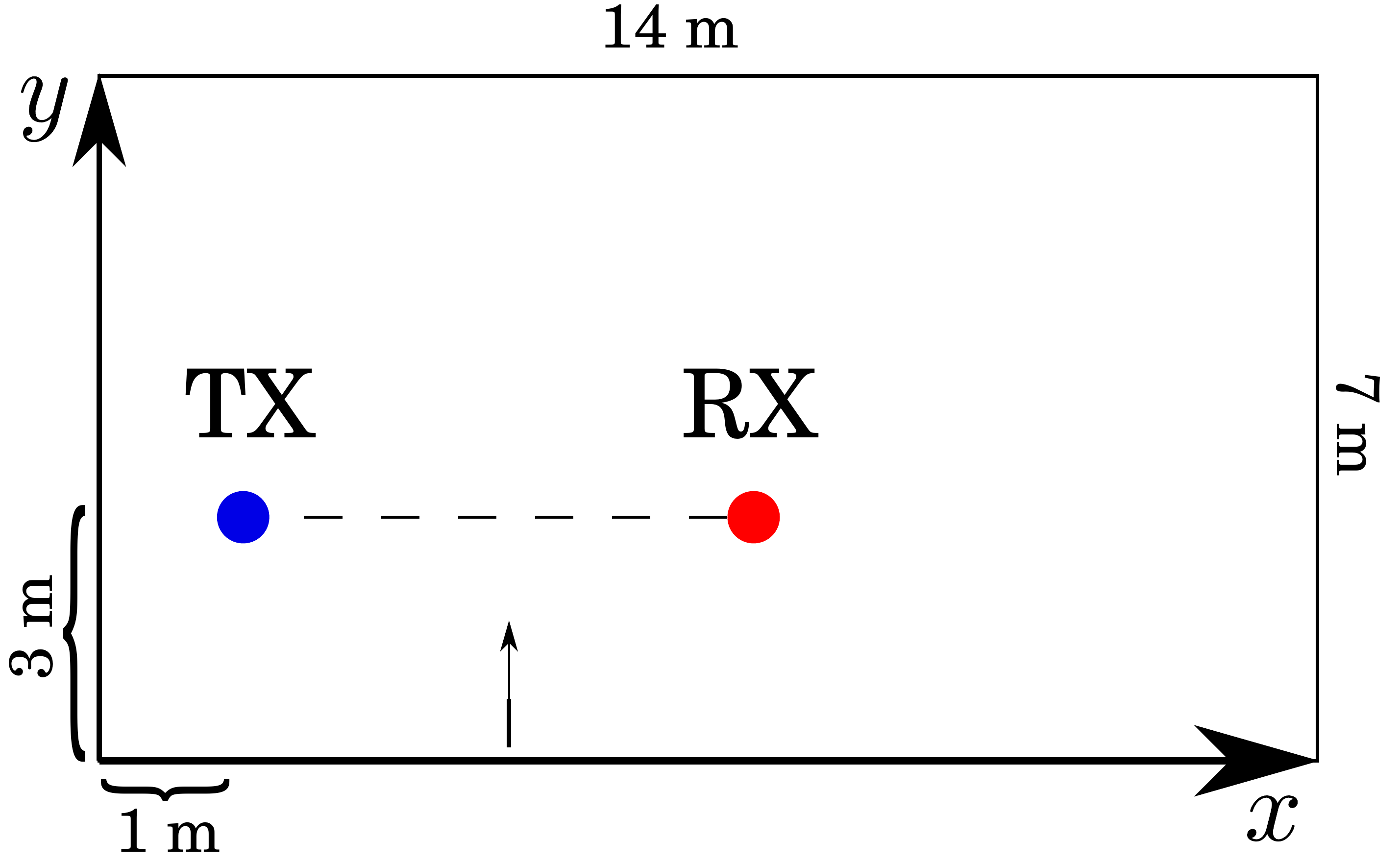}
		\vskip 4.5mm
		\caption{Static scenario.}
		\label{fig:scenario_hp}
	\end{subfigure}
	\hspace*{\fill}
	\begin{subfigure}[b]{.6\linewidth}
		\setlength\fwidth{\columnwidth}
		\setlength\fheight{\dfheight}
		\input{img/HumanPresence_refl2_60_GHz.tex}
		\caption{Max reflection order: 2}
		\label{fig:human_presence_refl2}
	\end{subfigure}
	\hspace*{\fill}
	
	\caption{Static scenario including second-order reflections.}
	\label{fig:human_presence}
\end{figure*}

In this section, we present a set of results obtained using the \textit{Blockage Manager}, to showcase the framework and offer some insight on the impact of blockage modeling on network simulations, and to compare the different diffraction models.
First, in \cref{sub:model_comparison}, we will show their general behavior as directly described by the equations, thus considering the effect that the obstacle has when passing through the direct ray between transmitter and receiver.

Then, \cref{sub:static_scenario} will consider a more realistic \textit{static} scenario, where transmitter and receiver are still fixed but placed in a room, and reflections, computed using a ray-tracer, also interact with the moving obstacle.

Finally, in \cref{sub:dynamic_scenario} a more complex \textit{dynamic} scenario is evaluated, where we consider an access point placed on the ceiling of a room and a user moving away from it.
In this case, multiple obstacles move in the scenario, to make the simulation even more realistic.

All the scenarios but one consider the obstacle(s) passing through the \gls{los} ray to better highlight the effect of the diffraction on the received power.
As detailed in the following, we consider also the case where the obstacles do not intercept the main ray, but only the reflected ones, to show that diffraction is relevant also in this situation.
For this work, we consider an (orthogonal-rectangular) thin screen, which provides a simplified yet realistic representation of the human body, according to measurements~\cite{kunisch08dked,gentile2021HumanPresence} and standards~\cite{itu-r-p526,metisChannelModels_D1.4}.
The obstruction loss is set to 10~dB, obtained by averaging the mean loss of the considered diffraction models.

\begin{table}[b]
  \centering
    \caption{Simulation parameters}
  \label{tab:params}
    \begin{tabular}{cc|cc}
        \toprule
        TX Antenna  &  8$\times$8 & RX Antenna  & 4$\times$4\\
        Antenna Spacing & $\frac{\lambda}{2}$  & Antenna Element & omni-directional\\
        Carrier Frequency &  60 GHz & Bandwidth & 2.16 GHz\\
        TX Power & 20 dBm & Noise Figure & 10 dB\\
        \bottomrule
    \end{tabular}
\end{table}

\begin{figure}[t!]
	\centering
	\setlength\fwidth{.7\columnwidth}
	\input{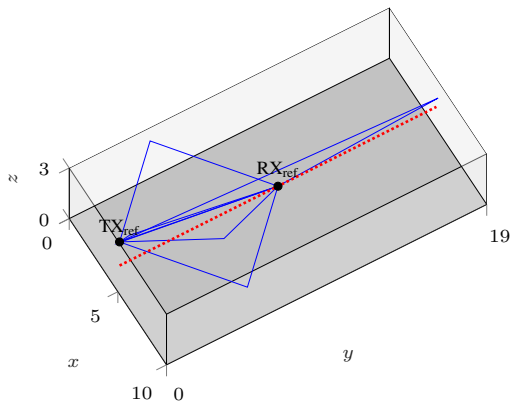}
	\caption{Dynamic scenario}
	\label{fig:scenario_ind1}
\end{figure}

First, the \textit{qd-realization} traces are imported in the \textit{Blockage Manager}, that applies the diffraction loss to the rays according to the obstacle configuration.
Link-layer simulations are then run on the traces using the same custom MATLAB simulator as in~\cite{testolina2019scalable,lecci2020accuracy}, to observe the blockage effect on the \gls{snr}.
However, given that the format of the output traces of the \textit{Blockage Manager} is the same as that of \textit{qd-realization}, they are compatible with other simulators such as \gls{ns3}~\cite{henderson2008network}.
The main parameters used for link-level simulations are listed in \cref{tab:params}.
Beamforming at both the transmitter and the receiver is achieved using an antenna array with omnidirectional antenna elements and $\lambda/2$ element spacing.

\subsection{Model Comparison}
\label{sub:model_comparison}

\begin{figure*}[t!]
	\begin{subfigure}[b]{\linewidth}
		\centering
		\setlength\fwidth{\columnwidth}
		% This file was created by matlab2tikz.
%
%The latest updates can be retrieved from
%  http://www.mathworks.com/matlabcentral/fileexchange/22022-matlab2tikz-matlab2tikz
%where you can also make suggestions and rate matlab2tikz.
%
\definecolor{mycolor1}{rgb}{0.00000,0.44700,0.74100}%
\definecolor{mycolor2}{rgb}{0.92900,0.69400,0.12500}%
\definecolor{mycolor3}{rgb}{0.49400,0.18400,0.55600}%
\definecolor{mycolor4}{rgb}{0.46600,0.67400,0.18800}%
\definecolor{mycolor5}{rgb}{0.30100,0.74500,0.93300}%
\begin{tikzpicture}
\pgfplotsset{every tick label/.append style={font=\scriptsize}}

\begin{axis}[%
width=0,
height=0,
at={(0,0)},
scale only axis,
xmin=0,
xmax=0,
xtick={},
ymin=0,
ymax=0,
ytick={},
axis background/.style={fill=white},
legend style={legend cell align=center,
              align=center,
              draw=white!15!black,
              at={(0, 0)},
              anchor=center,
              /tikz/every even column/.append style={column sep=2em}},
legend columns=3,
]
\addplot [color=mycolor1, line width=1.5pt]
  table[row sep=crcr]{%
0	0\\
};
\addlegendentry{METIS}

\addplot [color=black, line width=1.5pt]
  table[row sep=crcr]{%
  0	0\\
  };
\addlegendentry{Baseline}

\addplot [color=mycolor2, line width=1.5pt]
  table[row sep=crcr]{%
  0	0\\
  };
\addlegendentry{DKED+PC}

\addplot [color=mycolor3, dashed, line width=1.5pt]
  table[row sep=crcr]{%
  0	0\\
  };
\addlegendentry{DKED}

\addplot [color=mycolor4, line width=1.5pt]
  table[row sep=crcr]{%
  0	0\\
  };
\addlegendentry{ITU SE}

\addplot [color=mycolor5, line width=1.5pt]
  table[row sep=crcr]{%
  0	0\\
  };
\addlegendentry{Obstruction}

\end{axis}
\end{tikzpicture}%
	\end{subfigure}
	\vskip 0.5cm
	\hspace*{\fill}
	\begin{subfigure}[b]{\tfwidth}
		\centering
		\includegraphics[width=\columnwidth]{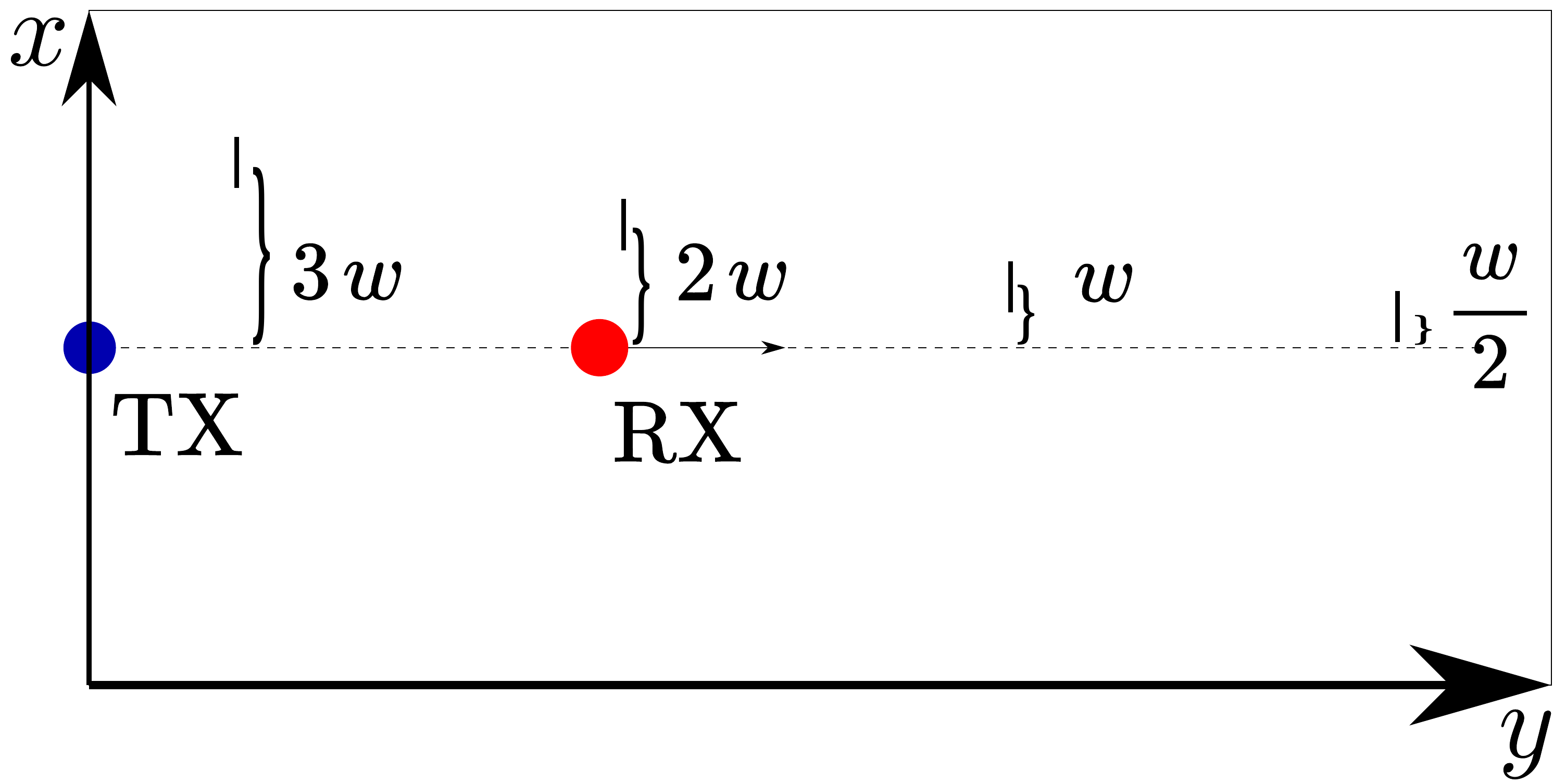}
		\vskip 5.7mm
		\caption{Obstacle locations (static).}
		\label{fig:indoor1_4_obs}
	\end{subfigure}
	\hspace*{\fill}
	\begin{subfigure}[b]{\tfwidth}
		\centering
		\setlength\fwidth{\columnwidth}
		\setlength\fheight{\dfheight}
		\input{img/Journal1Indoor1_static_4obs_60_GHz.tex}
		\caption{Full scenario}
		\label{fig:indoor1_observers_plot}
	\end{subfigure}
	\hspace*{\fill}
	\begin{subfigure}[b]{\tfwidth}
		\centering
		\setlength\fwidth{\columnwidth}
		\setlength\fheight{\dfheight}
		\input{img/Journal1Indoor1_static_4obs_60_GHz_zoom.tex}
		\caption{Zoom}
		\label{fig:indoor1_observers_zoom}
	\end{subfigure}
	\hspace*{\fill}
	
	\caption{Dynamic scenario with 4 static obstacles.}
	\label{fig:indoor1_observers}
\end{figure*}

\begin{figure*}[t!]
	\hspace*{\fill}
	\begin{subfigure}[b]{\tfwidth}
		\centering
		\includegraphics[width=\columnwidth]{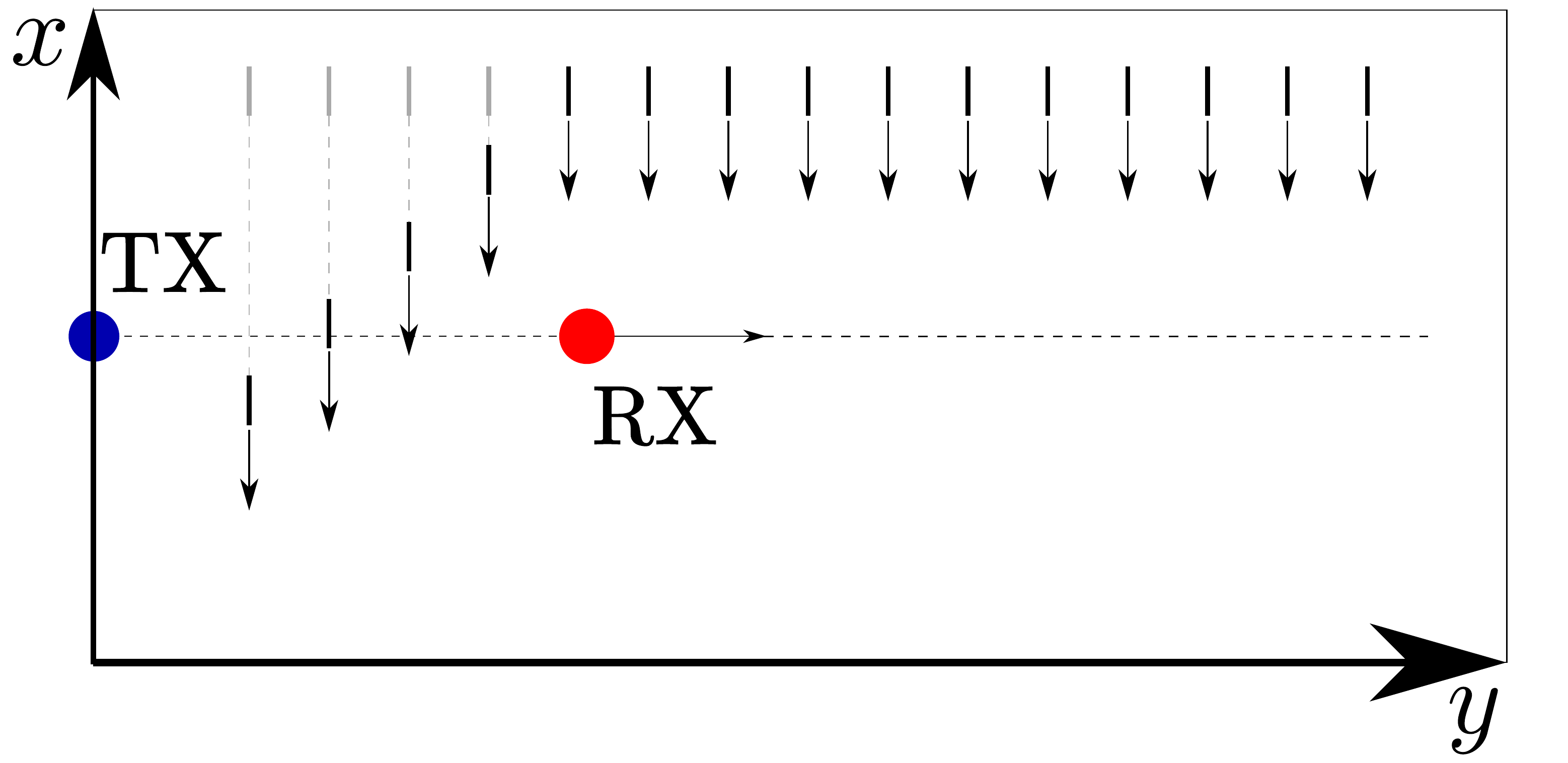}
		\vskip -1.5mm
		\caption{The 15 moving obstacles intercept the \gls{los} ray as the \gls{rx} moves through the room.}
		\label{fig:indoor1_15_obs}
	\end{subfigure}
	\hspace*{\fill}
	\begin{subfigure}[b]{\tfwidth}
		\centering
		\setlength\fwidth{\columnwidth}
		\setlength\fheight{\dfheight}
		\input{img/Journal1Indoor1_crossing_los_15obs_60_GHz.tex}
		\caption{Full scenario}
		\label{fig:indoor1_crossing_plot}
	\end{subfigure}
	\hspace*{\fill}
	\begin{subfigure}[b]{\tfwidth}
		\centering
		\setlength\fwidth{\columnwidth}
		\setlength\fheight{\dfheight}
		\input{img/Journal1Indoor1_crossing_los_15obs_60_GHz_zoom.tex}
		\caption{Zoom}
		\label{fig:indoor1_crossing_zoom}
	\end{subfigure}
	\hspace*{\fill}
	
	\caption{Dynamic scenario with 15 dynamic obstacles.}
	\label{fig:indoor1_crossing}
\end{figure*}

All implemented models have their own distinct characteristics derived by their formulations.
To give the reader an idea of their behavior, we show a set of key comparisons for these models in scenarios of interest.

We first consider a \gls{tx} at 1.6~m height, and a \gls{rx} at the same height placed 8~m away from the transmitter.
An obstacle of size 0.2$\times$1.7~m passes perpendicularly halfway between TX and RX, blocking the \gls{los} between the two nodes.
To highlight the effect of the models, we only process the direct ray, normalizing the received power to obtain the diffraction loss. %, and thus heights, widths, and distances involved are only accounted for how the diffraction acts on it \at{?}.
\cref{fig:models_60ghz} shows the losses of the different models at a 60~GHz carrier frequency.

In particular, the difference between the simple obstruction and the more complex diffraction models can be clearly observed.
Complex physics-based models, in fact, show an oscillatory behavior before and after the obstacle even intercepts the \gls{los}, due to its effect on the surrounding propagation environment, with peaks of almost 2~dB.
Furthermore, within the blockage region, very sharp deep fades affect the channel with over 10~dB extra losses, due to destructive interference among the rays curving around the obstacle.
This phenomenon, together with the high obstruction loss, can make channel estimation and adaptation harder, thus further reducing the communication efficiency in the presence of obstacles.
It is also possible to observe that different models have different average obstruction losses, making us question which one, if any, is close to the real-world measurements.

Instead, \cref{fig:models_obstacle_distance} shows what happens for the different models when the same obstacle moves starting close to the \gls{tx} and moving towards the RX, always obstructing the direct path between the two.
For all diffraction models, the distance between the obstacle and the nodes is taken into account when modeling its effect on the channel.
The figure shows a symmetric behavior, justified by the symmetry of wireless propagation, and all curves follow approximately the same trend.
Specifically, the loss tends to be higher when the obstacle is close to one of the nodes, and is lowest when the obstacle is exactly in the middle, a trend that is observed irrespective of the distance between TX and RX.
Intuitively, in fact, rays that need a sharper turn to surpass the obstacle (i.e., when the obstacle is close to one of the nodes) lose more energy than rays that need a shallower turn (i.e., when the obstacle is farthest).
A different explanation can be given by thinking about the apparent size of the obstacle, as seen from one of the nodes.
The same obstacle has a different apparent size depending on how far it is from its observer, where closer objects appear larger than farther objects and thus result in a deeper shadow.

Finally, it is important to remember that these models heavily depend on the carrier frequency.
In fact, in general higher frequencies will present sharper and larger losses, due to the larger electrical size of 
the obstacle with respect to the wavelength.
The short wavelength will also be more prone to creating constructive and destructive interference between the rays bending around the obstacle, creating deeper and more frequent loss peaks. 
Longer wavelength, instead, will be able to more easily bend around an obstacle, greatly reducing the obstruction losses.
These results can also be observed in \cref{fig:model_comparison_frequency}.

\subsection{Static Scenario}
\label{sub:static_scenario}

We tested our framework in a scenario inspired by~\cite{gentile2021HumanPresence}.
Specifically, two static nodes are placed in positions $p_1=(1, 3, 1.6)$ and $p_2=(9, 3, 1.6)$ within a 14$\times$7$\times$3~$\text{m}^3$ room, as shown in \cref{fig:scenario_hp}.
We simulated the channel with \textit{qd-realization}~\cite{qd-realization}, an open-source ray-tracing software for mmWave propagation, considering up to second-order reflections.

We then imported the channel trace into the \textit{Blockage Manager} software, configuring an orthogonal-rectangular screen moving from $p_{\rm start}=(5, 0, 0)$ upwards at 1.2~m/s, and sampled the channel every 3.4~ms for 1500~samples, for a total of about 5~s of simulation time.
We considered a thin screen with a width of 20~cm and a height of 1.7~m, in an attempt to emulate the size of the human body.

\cref{fig:human_presence_refl2} reports the \gls{snr} observed by the receiving node.
As the obstacle crosses through the \gls{los}, the \gls{snr} decreases rapidly, showing a behavior similar to that described in \cref{sub:model_comparison} both in shape and in amplitude.
Specifically, observing a diffraction loss so close to the one presented in \cref{fig:model_comparison} highlights how the \gls{los} ray carries the great majority of the signal power to the receiver, while the reflected ray can not compensate for the diffraction loss caused by the blocker.
On the other hand, while the average loss is similar to the one described in \cref{sub:model_comparison}, the deepest fades are mitigated by secondary reflected rays carrying far less power.
Namely, the troughs observed in most of the diffraction models often overestimate the fade depth~\cite{gentile2021HumanPresence}.
However, a complete channel simulation, where the transceivers and the obstacles are immersed in a realistic three-dimensional environment, tends to average them out, reducing the overall overshooting.

\subsection{Dynamic Scenario}
\label{sub:dynamic_scenario}

For this scenario, a different room is considered: the length and width are respectively equal to 19~m and 10~m, whereas the height is 3~m as described in \cref{fig:scenario_ind1}.

Although the environment, a rectangular room, is similar to the previous one, in this case the receiver is not fixed and the two nodes are not at the same height.
Indeed the transmitter has a fixed position in $p_1 = (5,0.1,2.9)$ and the receiver starts from a position right below the transmitter, $p_{\text{start}} = (5,0.1,1.5)$, and moves away from it at 1.2~m/s for about 15.7~s, reaching the position $p_{\text{final}} = (5,18.9,1.5)$ next to the farthest wall from the \gls{tx} position.
The total number of samples is 3133, achieved using a sampling period of 5~ms.
Following the results from~\cite{lecci2020accuracy}, we simulated up to second-order reflections, and excluded rays less powerful than the most prominent one by more than 40~dB.
These parameters were shown to yield a good balance between computational effort and accuracy of results.
We then imported the results into the \textit{Blockage Manager} software with two different obstacle settings.

In the first scenario, reported in \cref{fig:indoor1_4_obs}, 4 static obstacles of size 0.4$\times$1.7~$\text{m}^2$ are considered, none of them blocking the direct ray.
The obstacles are placed at decreasing distances from the \gls{los}, specifically in $p_1=(5.2,15.14,0)$, $p_2=(5.8,7.62,0)$, $p_3=(5.4,11.38,0)$, and $p_4=(6.2,3.86,0)$.

As shown in \cref{fig:indoor1_observers}, although the obstacles do not create strong or sudden effects on the received power, they still play an important role in the total received power, showing differences with respect to the baseline of up to 8.2~dB.
On the other hand, as the \gls{rx} passes by the first three obstacles, all models behave almost indistinguishably from each other, with an absolute error of at most 1.6~dB with respect to the simple obstruction model.
On the contrary, the diffraction introduced by the last and closest screen is significant, with the \gls{snr} of the diffraction models presenting a difference of about 5~dB from that of the obstruction model.
This justifies our design choice of setting a distance threshold on the diffraction model, so that obstacles far enough from the ray will not affect it, making it sufficient to only model the most significant effects of the diffraction.
Similarly, we also allow the user to fall back to a simpler obstruction model for secondary or less powerful rays, which have a much smaller effect on the total received power.

Finally, in the second scenario we consider 15 obstacles traversing the room at regular intervals, as represented in \cref{fig:indoor1_15_obs}.
The results in \cref{fig:indoor1_crossing} show all the effects discussed so far in the previous sections, combined in a more complex and realistic scenario.
In particular, it is possible to observe that (i) all models behave as expected during obstruction (highlighted in gray), (ii) obstacles affect the channel also when not directly blocking the \gls{los}, (iii) constructive interference before and after obstruction can actually be very significant, even overshooting the baseline, (iv) using a constant obstruction loss does not represent well the complex interaction between transceivers and obstacles, (v) when the receiver reaches the other side of the room, and with multiple rays having similar power to the direct path, obstruction is less severe and small scale fading is actually the main concern.

\section{Conclusions}
\label{sec:conclusions}
In this paper we presented a novel open source tool, the \textit{Blockage Manager}, to model the diffraction by dynamic obstacles in \gls{rt} traces.
Namely, the software post-processes the output from a ray-tracer, allowing the user to introduce an arbitrary number of obstacles in the simulation, and modeling their interactions with the rays choosing from a number of diffraction models.
The potential of the software was showcased with simple yet insightful network simulations, which already provided interesting results.
We hope that this tool lays the foundations for more accurate, large-scale studies on the effects of blockage on \gls{mmwave} networks.

Comparison with real-world data is an essential step to validate the framework.
A more precise calibration of the diffraction models against measurements is currently being considered, and we plan to provide a calibration toolbox as an additional module for the framework.

Future steps will also focus on making the \textit{Blockage Manager} able to describe a much wider variety of scenarios, introducing new obstacles and diffraction models.
For instance, a foliage model could be implemented to increase the accuracy in outdoor environments where the signal propagates through trees and vegetation.
Furthermore, \gls{utd} may be considered as a more general approach to model the diffraction, including obstacles of arbitrary shape and dimension.

Finally, a full stack analysis of a real scenario with accurate obstacle models is essential to understand more precisely the actual effect of multiple, dynamic obstacles on the final user experience in different practical situations.

\bibliographystyle{IEEEtran}
\bibliography{../bibl} 

% Generated by IEEEtran.bst, version: 1.14 (2015/08/26)
\begin{thebibliography}{10}
\providecommand{\url}[1]{#1}
\csname url@samestyle\endcsname
\providecommand{\newblock}{\relax}
\providecommand{\bibinfo}[2]{#2}
\providecommand{\BIBentrySTDinterwordspacing}{\spaceskip=0pt\relax}
\providecommand{\BIBentryALTinterwordstretchfactor}{4}
\providecommand{\BIBentryALTinterwordspacing}{\spaceskip=\fontdimen2\font plus
\BIBentryALTinterwordstretchfactor\fontdimen3\font minus
  \fontdimen4\font\relax}
\providecommand{\BIBforeignlanguage}[2]{{%
\expandafter\ifx\csname l@#1\endcsname\relax
\typeout{** WARNING: IEEEtran.bst: No hyphenation pattern has been}%
\typeout{** loaded for the language `#1'. Using the pattern for}%
\typeout{** the default language instead.}%
\else
\language=\csname l@#1\endcsname
\fi
#2}}
\providecommand{\BIBdecl}{\relax}
\BIBdecl

\bibitem{rappaport2013millimeter}
T.~S. Rappaport, S.~Sun, R.~Mayzus, H.~Zhao, Y.~Azar, K.~Wang, G.~N. Wong,
  J.~K. Schulz, M.~Samimi, and F.~Gutierrez, ``Millimeter wave mobile
  communications for {5G} cellular: It will work!'' \emph{IEEE Access}, vol.~1,
  pp. 335--349, 2013.

\bibitem{rangan2017potentials}
S.~Rangan, T.~S. Rappaport, and E.~Erkip, ``Millimeter-wave cellular wireless
  networks: Potentials and challenges,'' \emph{Proceedings of the IEEE}, vol.
  102, no.~3, pp. 366--385, Mar. 2014.

\bibitem{dengMmwaveDiffraction}
S.~{Deng}, G.~R. {MacCartney}, and T.~S. {Rappaport}, ``Indoor and outdoor {5G}
  diffraction measurements and models at 10, 20, and 26 {GHz},'' in \emph{IEEE
  Global Communications Conference (GLOBECOM)}, Dec. 2016.

\bibitem{hemadeh2018millimeter}
I.~A. {Hemadeh}, K.~{Satyanarayana}, M.~{El-Hajjar}, and L.~{Hanzo},
  ``Millimeter-wave communications: Physical channel models, design
  considerations, antenna constructions, and link-budget,'' \emph{{IEEE}
  Communications Surveys and Tutorials}, vol.~20, no.~2, pp. 870--913, Second
  Quarter 2018.

\bibitem{testolina2020simplified}
M.~Lecci, P.~Testolina, M.~Giordani, M.~Polese, T.~Ropitault, C.~Gentile,
  N.~Varshney, A.~Bodi, and M.~Zorzi, ``{Simplified Ray Tracing for the
  Millimeter Wave Channel: A Performance Evaluation},'' in \emph{Information
  Theory and Applications Workshop (ITA)}, San Diego, CA, USA, Feb. 2020.

\bibitem{lecci2020accuracy}
M.~Lecci, P.~Testolina, M.~Polese, M.~Giordani, and M.~Zorzi, ``{Accuracy
  Versus Complexity for mmWave Ray-Tracing: A Full Stack Perspective},''
  \emph{IEEE Transactions on Wireless Communications}, vol.~20, no.~12, pp.
  7826--7841, Dec. 2021.

\bibitem{gentile2021HumanPresence}
A.~Bhardwaj, D.~Caudill, C.~Gentile, J.~Chuang, J.~Senic, and D.~G. Michelson,
  ``Geometrical-empirical channel propagation model for human presence at 60
  {GHz},'' \emph{IEEE Access}, vol.~9, pp. 38\,467--38\,478, 2021.

\bibitem{macCartney2016humanBlockage}
G.~R. {MacCartney}, S.~{Deng}, S.~{Sun}, and T.~S. {Rappaport},
  ``Millimeter-wave human blockage at 73 {GHz} with a simple double knife-edge
  diffraction model and extension for directional antennas,'' in \emph{IEEE
  84th Vehicular Technology Conference (VTC-Fall)}, Sep. 2016, pp. 1--6.

\bibitem{qd-realization}
{wigig-tools}, ``qd-realization,''
  \url{https://github.com/signetlabdei/qd-realization/tree/feature/power-threshold},
  open-source ray-tracer implementation.

\bibitem{bai2015coverage}
T.~Bai and R.~W. Heath, ``Coverage and rate analysis for millimeter-wave
  cellular networks,'' \emph{IEEE Transactions on Wireless Communications},
  vol.~14, no.~2, pp. 1100--1114, Feb. 2015.

\bibitem{andrews2017modeling}
J.~G. {Andrews}, T.~{Bai}, M.~N. {Kulkarni}, A.~{Alkhateeb}, A.~K. {Gupta}, and
  R.~W. {Heath}, ``Modeling and analyzing millimeter wave cellular systems,''
  \emph{IEEE Transactions on Communications}, vol.~65, no.~1, pp. 403--430,
  Jan. 2017.

\bibitem{ferrand2016trends}
P.~Ferrand, M.~Amara, S.~Valentin, and M.~Guillaud, ``Trends and challenges in
  wireless channel modeling for evolving radio access,'' \emph{{IEEE}
  Communications Magazine}, vol.~54, no.~7, pp. 93--99, Jul. 2016.

\bibitem{3gpp.38.901}
3GPP, ``Study on channel model for frequencies from 0.5 to 100 {GHz},'' 3rd
  Generation Partnership Project (3GPP), Technical Report (TR) 38.901, Jun.
  2018, version 15.0.0.

\bibitem{gapeyenko2018analytical}
\BIBentryALTinterwordspacing
M.~Gapeyenko, V.~Petrov, D.~Moltchanov, S.~Andreev, Y.~Koucheryavy, M.~Valkama,
  M.~R. Akdeniz, and N.~Himayat, ``{An Analytical Representation of the 3GPP 3D
  Channel Model Parameters for MmWave Bands},'' in \emph{Proceedings of the 2nd
  ACM Workshop on Millimeter Wave Networks and Sensing Systems}, ser. mmNets
  ’18.\hskip 1em plus 0.5em minus 0.4em\relax New Delhi, India: Association
  for Computing Machinery, 2018, p. 33–38. [Online]. Available:
  \url{https://doi.org/10.1145/3264492.3264498}
\BIBentrySTDinterwordspacing

\bibitem{zugno2020implementation}
T.~Zugno, M.~Polese, N.~Patriciello, B.~Bojovi\'{c}, S.~Lagen, and M.~Zorzi,
  ``{Implementation of a Spatial Channel Model for Ns-3},'' in \emph{ACM
  Workshop on Ns-3 (WNS3)}, Gaithersburg, MD, USA, Jun. 2020.

\bibitem{lecci2020qd}
M.~Lecci, M.~Polese, C.~Lai, J.~Wang, C.~Gentile, N.~Golmie, and M.~Zorzi,
  ``{Quasi-Deterministic Channel Model for mmWaves: Mathematical Formalization
  and Validation},'' in \emph{IEEE Global Communications Conference
  (GLOBECOM)}, Taipei, Taiwan, Dec. 2020.

\bibitem{degliesposti2014rt}
V.~{Degli-Esposti}, F.~{Fuschini}, E.~M. {Vitucci}, M.~{Barbiroli}, M.~{Zoli},
  L.~{Tian}, X.~{Yin}, D.~A. {Dupleich}, R.~{Müller}, C.~{Schneider}, and
  R.~S. {Thomä}, ``Ray-tracing-based {mm-Wave} beamforming assessment,''
  \emph{IEEE Access}, vol.~2, pp. 1314--1325, 2014.

\bibitem{mckown91rt}
J.~McKown and R.~Hamilton, ``Ray tracing as a design tool for radio networks,''
  \emph{IEEE Network}, vol.~5, no.~6, pp. 27--30, Nov. 1991.

\bibitem{schuler08scatteringCenters}
K.~Schuler, D.~Becker, and W.~Wiesbeck, ``Extraction of virtual scattering
  centers of vehicles by ray-tracing simulations,'' \emph{IEEE Transactions on
  Antennas and Propagation}, vol.~56, no.~11, pp. 3543--3551, Nov. 2008.

\bibitem{ghaddar07conductingCylinder}
M.~Ghaddar, L.~Talbi, T.~A. Denidni, and A.~Sebak, ``A conducting cylinder for
  modeling human body presence in indoor propagation channel,'' \emph{IEEE
  Transactions on Antennas and Propagation}, vol.~55, no.~11, pp. 3099--3103,
  Nov. 2007.

\bibitem{metisChannelModels_D1.4}
``{METIS Channel Models},'' Deliverable D1.4, Feb. 2015.

\bibitem{medbo2013channel}
J.~Medbo and F.~Harrysson, ``Channel modeling for the stationary {UE}
  scenario,'' in \emph{7th European Conference on Antennas and Propagation
  (EuCAP)}, Gothenburg, Sweden, Apr. 2013.

\bibitem{casciato01thesis}
M.~D. Casciato, ``Radio wave diffraction and scattering models for wireless
  channel simulation,'' Ph.D. dissertation, University of Michigan, 2001.

\bibitem{prado21enhance}
D.~Prado-Alvarez, S.~Inca, D.~Martín-Sacristán, and J.~F. Monserrat,
  ``Millimeter-wave human blockage model enhancements for directional antennas
  and multiple blockers,'' \emph{IEEE Communications Letters}, vol.~25, no.~9,
  pp. 2776--2780, Sep. 2021.

\bibitem{slezak18under}
C.~Slezak, M.~Zhang, M.~Mezzavilla, and S.~Rangan, ``{Understanding End-to-End
  Effects of Channel Dynamics in Millimeter Wave 5G New Radio},'' in \emph{IEEE
  19th International Workshop on Signal Processing Advances in Wireless
  Communications (SPAWC)}, 2018, pp. 1--5.

\bibitem{jacob11humanBlockage}
M.~Jacob, S.~Priebe, A.~Maltsev, A.~Lomayev, V.~Erceg, and T.~Kurner, ``{A ray
  tracing based stochastic human blockage model for the IEEE 802.11ad 60 GHz
  channel model},'' in \emph{5th European Conference on Antennas and
  Propagation (EuCAP)}, Rome, Italy, Apr. 2011.

\bibitem{tgad_channel_model}
A.~Maltsev \emph{et~al.}, ``Channel models for 60 {GHz} {WLAN} systems,''
  09/0334r8, IEEE Task Group ad (TGad), Tech. Rep., May 2010.

\bibitem{zhang2019will}
M.~{Zhang}, M.~{Polese}, M.~{Mezzavilla}, J.~{Zhu}, S.~{Rangan}, S.~{Panwar},
  and M.~{Zorzi}, ``{Will TCP Work in mmWave 5G Cellular Networks?}''
  \emph{{IEEE} Communications Magazine}, vol.~57, no.~1, pp. 65--71, January
  2019.

\bibitem{polese17mobilitymanagement}
M.~Polese, M.~Mezzavilla, S.~Rangan, and M.~Zorzi, ``{Mobility Management for
  TCP in MmWave Networks},'' in \emph{1st ACM Workshop on Millimeter-Wave
  Networks and Sensing Systems (mmNets)}, ser. mmNets '17, Snowbird, Utah, USA,
  Oct. 2017.

\bibitem{polese17milliproxy}
M.~Polese, M.~Mezzavilla, M.~Zhang, J.~Zhu, S.~Rangan, S.~Panwar, and M.~Zorzi,
  ``{milliProxy: A TCP proxy architecture for 5G mmWave cellular systems},'' in
  \emph{51st Asilomar Conference on Signals, Systems, and Computers}, 2017, pp.
  951--957.

\bibitem{polese17tcpin5g}
M.~Polese, R.~Jana, and M.~Zorzi, ``{TCP in 5G mmWave networks: Link level
  retransmissions and MP-TCP},'' in \emph{IEEE Conference on Computer
  Communications Workshops (INFOCOM WKSHPS)}, 2017, pp. 343--348.

\bibitem{polese17tcp}
------, ``{TCP and MP-TCP in 5G mmWave Networks},'' \emph{IEEE Internet
  Computing}, vol.~21, no.~5, pp. 12--19, Sep. 2017.

\bibitem{mezzavilla2018end}
M.~{Mezzavilla}, M.~{Zhang}, M.~{Polese}, R.~{Ford}, S.~{Dutta}, S.~{Rangan},
  and M.~{Zorzi}, ``End-to-end simulation of {5G} {mmWave} networks,''
  \emph{{IEEE} Communications Surveys and Tutorials}, vol.~20, no.~3, pp.
  2237--2263, Third Quarter 2018.

\bibitem{ren2021tcpMmwave}
Y.~Ren, W.~Yang, X.~Zhou, H.~Chen, and B.~Liu, ``{A survey on TCP over
  mmWave},'' \emph{Computer Communications}, vol. 171, pp. 80--88, Apr. 2021.

\bibitem{Imran22mptcpSmartphones}
I.~Khan, M.~Ghoshal, S.~Aggarwal, D.~Koutsonikolas, and J.~Widmer, ``{Multipath
  TCP in Smartphones Equipped with Millimeter Wave Radios},'' in \emph{15th ACM
  Workshop on Wireless Network Testbeds, Experimental Evaluation and
  Characterization (WiNTECH)}, New Orleans, LA, USA, Jan. 2022.

\bibitem{lee19mptcp}
C.~Lee, S.~Song, H.~Cho, G.~Lim, and J.-M. Chung, ``{Optimal Multipath TCP
  Offloading Over 5G NR and LTE Networks},'' \emph{IEEE Wireless Communications
  Letters}, vol.~8, no.~1, pp. 293--296, Feb. 2019.

\bibitem{simpy}
\BIBentryALTinterwordspacing
A.~Meurer, C.~P. Smith, M.~Paprocki, O.~\v{C}ert\'{i}k, S.~B. Kirpichev,
  M.~Rocklin, A.~Kumar, S.~Ivanov, J.~K. Moore, S.~Singh, T.~Rathnayake,
  S.~Vig, B.~E. Granger, R.~P. Muller, F.~Bonazzi, H.~Gupta, S.~Vats,
  F.~Johansson, F.~Pedregosa, M.~J. Curry, A.~R. Terrel, v.~Rou\v{c}ka,
  A.~Saboo, I.~Fernando, S.~Kulal, R.~Cimrman, and A.~Scopatz, ``Sympy:
  symbolic computing in python,'' \emph{PeerJ Computer Science}, vol.~3, p.
  e103, Jan. 2017. [Online]. Available:
  \url{https://doi.org/10.7717/peerj-cs.103}
\BIBentrySTDinterwordspacing

\bibitem{shapely}
\BIBentryALTinterwordspacing
S.~Gillies \emph{et~al.}, ``Shapely: manipulation and analysis of geometric
  objects,'' 2007--. [Online]. Available:
  \url{https://github.com/shapely/shapely}
\BIBentrySTDinterwordspacing

\bibitem{itu-r-p526}
ITU-R, ``Propagation by diffraction,'' Recommendation ITU-R P.526-15, Oct.
  2019.

\bibitem{ford16ns3mmwave}
R.~Ford, M.~Zhang, S.~Dutta, M.~Mezzavilla, S.~Rangan, and M.~Zorzi, ``{A
  Framework for End-to-End Evaluation of 5G MmWave Cellular Networks in
  Ns-3},'' in \emph{Proceedings of the ACM Workshop on Ns-3 (WNS3)}, Seattle,
  WA, USA, Jun. 2016.

\bibitem{assasa16implementation11ad}
H.~Assasa and J.~Widmer, ``{Implementation and Evaluation of a WLAN IEEE
  802.11ad Model in Ns-3},'' in \emph{Proceedings of the Workshop on Ns-3
  (WNS3)}, Seattle, WA, USA, Jun. 2016.

\bibitem{assasa17extending11ad}
------, ``{Extending the IEEE 802.11ad Model: Scheduled Access, Spatial Reuse,
  Clustering, and Relaying},'' in \emph{Proceedings of the Workshop on Ns-3
  (WNS3)}, Porto, Portugal, Jun. 2017.

\bibitem{kunisch08dked}
J.~Kunisch and J.~Pamp, ``Ultra-wideband double vertical knife-edge model for
  obstruction of a ray by a person,'' in \emph{IEEE International Conference on
  Ultra-Wideband}, Hannover, Germany, Sep. 2008.

\bibitem{testolina2019scalable}
P.~{Testolina}, M.~{Lecci}, M.~{Polese}, M.~{Giordani}, and M.~{Zorzi},
  ``Scalable and accurate modeling of the millimeter wave channel,'' in
  \emph{International Conference on Computing, Networking and Communications
  (ICNC)}, Feb. 2020, pp. 969--974.

\bibitem{henderson2008network}
T.~R. Henderson, M.~Lacage, G.~F. Riley, C.~Dowell, and J.~Kopena, ``Network
  simulations with the ns-3 simulator,'' vol.~14, no.~14, 2008, p. 527.

\end{thebibliography}

\end{document}